# Thematic agreement assessment of gridded, multi-modal geospatial datasets of different semantics and spatial granularities

Johannes H. Uhl[1,2], Stefan Leyk[1,2]

[1]Department of Geography, University of Colorado Boulder, Boulder, Colorado, USA
[2]Institute of Behavioral Science, University of Colorado Boulder, Boulder, Colorado, USA

*Corresponding author: Johannes.Uhl@colorado.edu

**Abstract**
This paper presents a method for thematic agreement assessment of geospatial data products of different semantics and spatial granularities, which may be affected by spatial offsets between test and reference data. The proposed method uses a multi-scale framework allowing for a probabilistic evaluation whether thematic disagreement between datasets is induced by spatial offsets due to different nature of the datasets or not. We test our method using real-estate derived settlement locations and remote-sensing derived building footprint data.

## I INTRODUCTION

Advances in geospatial data acquisition methods, geographic information extraction techniques, and the development of spatial data storage and analysis infrastructure catalyze the steadily increasing availability of multi-source, multi-sensor, and multi-modal spatial and spatio-temporal digital data. While such data richness offers unprecedented opportunities to better understand the measured geographic processes through meaningful integration of such data, it also causes increasing levels of discrepancies and inconsistencies among datasets, as well as varying levels of uncertainty across different datasets.

Geospatial data uncertainty is typically quantified by comparing the data under test against a reference dataset measuring the same underlying phenomenon at presumably higher accuracy (FGDC 1998). However, there are situations when such optimal reference data is not available. While several frameworks for multi-scale accuracy assessment have been proposed (e.g. Pontius 2002, Pontius and Suedmeyer 2004, Pontius et al. 2004, Pontius and Cheuk 2006, Pontius et al. 2008, Pontius et al. 2011, Stehman and Wickham 2011, Zhu et al. 2013, Yan et al. 2014), few approaches explicitly assess cross-scale accuracy trajectories (e.g., Uhl & Leyk 2017). Herein, we illustrate such a case, and present a method for cross-scale thematic agreement assessment when multi-modal datasets of related, but not identical semantics, and of different levels of spatial granularity are compared against each other. More specifically, we compare a spatially fine-grained surface of housing counts derived from parcel centroids and address points against a surface of remote-sensing derived building footprint counts.

## II DATA

The Zillow Transaction and Assessment Dataset (ZTRAX) (Zillow 2016) is an attribute-rich building stock database, compiled from heterogeneous data sources such as county-level tax assessment data, and contains information on land use, building age, material, and size for over 150 million parcels in the United States. While thematically rich, spatial information in that database is constrained to discrete locations, presumed to be inside the corresponding cadastral parcel units and thus abstract representations of approximate address points. Thus, spatial accuracy of the locational information is a function of parcel size, and thus, is expected to decrease from urban, residential areas (Fig. 1a) towards rural settlements, where cadastral parcels may be large (Fig. 1b). Based on these data, we generated multi-temporal, spatially fine-grained, gridded layers of housing counts, at a resolution of 250m (Historical Settlement Data Compilation for the US, HISDAC-US, Leyk & Uhl 2018, see Fig. 3a). To validate our data product at national scale, we use the Microsoft building footprint dataset, which has been derived automatically from aerial imagery and contains over 125 million building footprints in the US (Microsoft 2018), and thus, can be expected to be highly precise and complete. In fact, to the best of our knowledge, there is no other dataset of US-wide coverage that could be used as reference data for this scenario. For compatibility reasons and to keep computational cost to a minimum, we generated a building footprint count surface from these data, likewise at a spatial resolution of 250m to be used as reference data (Fig. 3b).

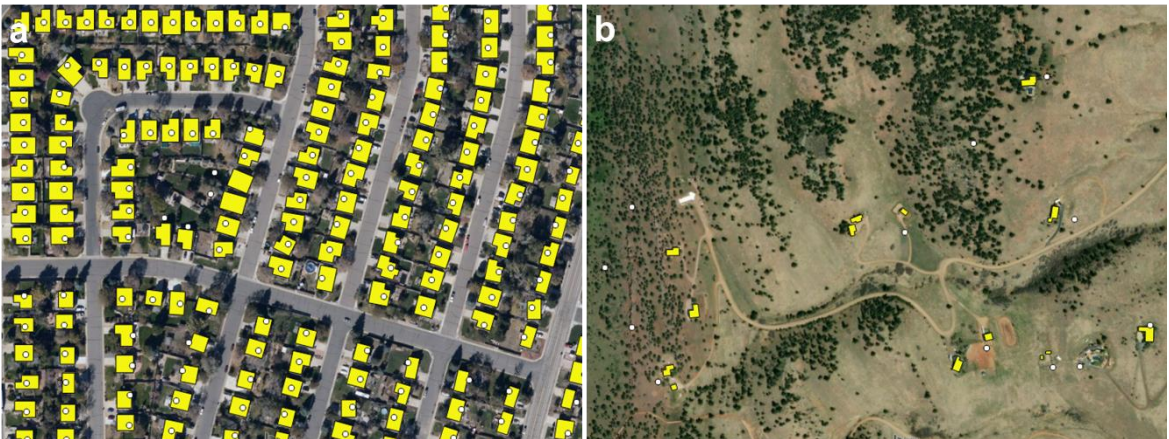

Figure 1: Data used for the thematic agreement assessment: ZTRAX data points (white) and building footprint vectors (yellow) in (a) dense urban areas, highly precise and well-aligned, and (b) in rural parts of the US, where we observe spatial offsets between ZTRAX point locations and building footprints.

## III METHOD

As illustrated in Fig. 1b), large rural cadastral parcels cause spatial offsets between the building footprint and the location recorded in ZTRAX. These offsets may be large in sparsely populated, rural regions, and thus, may result in spatial disagreement between the (rasterized) test and reference data, even though both datasets are in agreement (i.e., ZTRAX location and building footprint are within the same parcel boundaries). Hence, spatial disagreement (i.e., false positive or false negative instances) is regarded „true", if we can rule out that the disagreement is induced by spatial offsets due to different semantics (i.e., parcel centroid / address point vs. building footprint) and spatial granularity (i.e., discrete point vs. polygon) between underlying test and reference data. Our method models the probability that thematic disagreement is induced by such spatial offsets, and works as follows: First, classical map comparison is conducted for the (binarized) test and reference surfaces at original spatial resolution (here: 250m), resulting in a categorical raster dataset indicating the agreement type (true positives, true negatives, false positives, false negatives, i.e., $TP_{250}, TN_{250}, FP_{250}, FN_{250}$, respectively). Then, the binarized test and reference surfaces are downsampled by factor 2, and pixel-wise agreement types are re-computed (i.e., $TP_{500}, TN_{500}, FP_{500}, FN_{500}$). This is done iteratively for a specified number of downsampling factors (here: up to factor 4, corresponding to a pixel size of 2000m), which indicates the spatial range within which offsets as described above are assumed to occur. The agreement type surfaces of all downsampling levels are then upsampled to the native resolution (i.e., 250m) and stacked into a multi-scale data cube (see Fig. 2). Based on this cube, pixel-wise cross-scale trajectories are extracted for each pixel misclassified at native resolution (see Table 1). When a cross-scale trajectory switches from FP to TP, or from FN to TP, respectively, a probability of offset-induced misclassification is assigned to the pixel as a function of the aggregation level where this switch occurs. This probability is lowest for pixels that remain in FP or FN categories across all scales, and highest if the switch to TP occurs immediately after the first downsampling step.

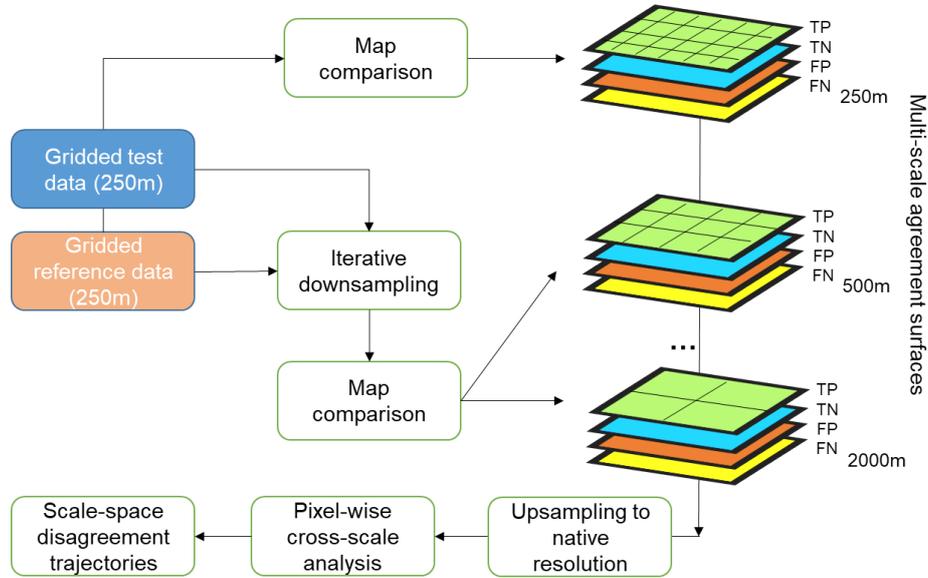

Figure 2: Processing workflow to generate the cross-scale disagreement composite surface. Source: Uhl et al. (2021).

Table 1: Cross-scale disagreement trajectories and assigned offset-induced misclassification probability. Source: Uhl et al. (2021).

| Spatial aggregation level | | | | Probability of offset-induced misclassification |
|---|---|---|---|---|
| 250m | 500m | 1000m | 2000m | |
| FP | FP | FP | FP | lowest |
| FP | FP | FP | *TP* | low |
| FP | FP | *TP* | *TP* | medium |
| FP | *TP* | *TP* | *TP* | highest |
| FN | FN | FN | FN | lowest |
| FN | FN | FN | *TP* | low |
| FN | FN | *TP* | *TP* | medium |
| FN | *TP* | *TP* | *TP* | highest |

A subset of resulting surfaces indicating FPs and FNs including their estimated offset-induced misclassification probability, as well as the TPs, is shown in Fig. 3.

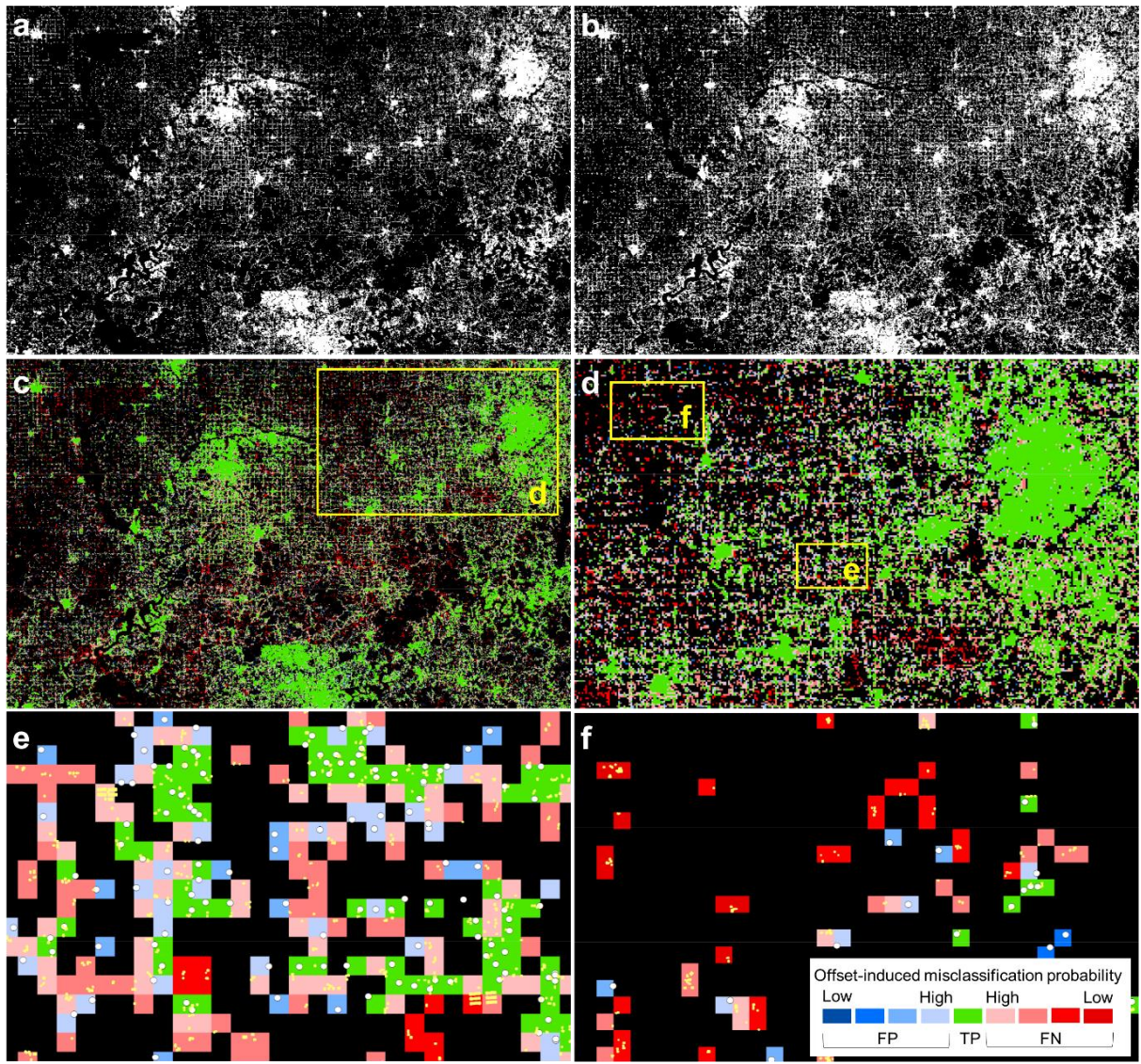

Figure 3: Illustrating the method for thematic agreement assessment under positional uncertainty. (a) Binarized, ZTRAX-derived settled areas (white), (b) corresponding reference surface derived from Microsoft building footprint data, (c) resulting spatial disagreement surface indicating the estimated offset-induced misclassification probability, (d) subset shown for a region west of Springfield, Missouri, and enlargements showing regions characterized by (e) disagreement likely introduced by spatial offsets, and (f) false negatives unlikely introduced by spatial offsets, including original ZTRAX locations (white dots) and Microsoft building footprints (yellow). Source: Uhl et al. (2021).

## IV CONCLUSIONS AND OUTLOOK

The presented method constitutes a computationally feasible method for map comparison if differences in semantics and spatial granularity between test and reference data cause spatial offsets between corresponding objects across both datasets. While traditional, rigorous map comparison at native resolution would overestimate spatial disagreement, the proposed method provides a framework for probabilistic analysis of misclassified instances. Future work will incorporate the generation of cross-scale accuracy measures, as well as the incorporation of built-up density estimates derived from the reference data, allowing for stratified assessment of the ZTRAX-derived settlement extents across the rural-urban continuum.

**ACKNOWLEDGEMENTS**

Funding for this work was provided through NSF's Humans, Disasters, and the Built Environment program (award #1924670 to CU Boulder), as well as the Eunice Kennedy Shriver National Institute of Child Health & Human Development of the National Institutes of Health under Award Number P2CHD066613. The content is solely the responsibility of the authors and does not necessarily represent the official views of the National Institutes of Health. We gratefully acknowledge access to the Zillow Transaction and Assessment Dataset (ZTRAX) through a data use agreement between the University of Colorado Boulder and Zillow Group, Inc. More information on accessing the data can be found at http://www.zillow.com/ztrax. The results and opinions are those of the author(s) and do not reflect the position of Zillow Group. Support by Zillow Group, Inc. is gratefully acknowledged.